\documentclass[aps,prd,10pt,a4paper,superscriptaddress,preprintnumbers,showpacs,notitlepage,nofootinbib]{revtex4-1}

\usepackage[english]{babel}
\usepackage{graphicx}
\usepackage{amsmath,amssymb}
\usepackage{xcolor}
\usepackage{eufrak}
\usepackage{caption}
\usepackage{subcaption}
\usepackage{hyperref}    
\usepackage{stmaryrd}
\usepackage{comment}

\def\sqr#1#2{{\vcenter{\vbox{\hrule height.#2pt\hbox{\vrule
width.#2pt height#1pt \kern#1pt\vrule width.#2pt}\hrule height.#2pt}}}}

\begin{document}

\title{Gravitational radiation from compact binary systems\\in Einstein-Maxwell-dilaton theories}
\author{F\'elix-Louis Juli\'e}

\affiliation{APC, Universit\'e Paris Diderot,\\
CNRS, CEA, Observatoire de Paris, Sorbonne Paris Cit\'e\\
 10, rue Alice Domon et L\'eonie Duquet, F-75205 Paris CEDEX 13, France.}

\date{September 13th, 2018}

\begin{abstract}
We derive the energy fluxes radiated by compact binary systems, including ``hairy" black holes, in Einstein-Maxwell-dilaton theories, for circular orbits and at quadrupolar order. This enables to include in their resummed effective-one-body (EOB) dynamics the effect of the radiation reaction force at the origin of their inspiral and merger. We also exhibit typical examples of the resulting tensor and scalar waveforms.
\end{abstract}

\maketitle


\vspace*{0.3cm}

The Einstein-Maxwell-dilaton theories consist in supplementing general relativity with massless scalar and vector fields, and are described by the following Einstein frame action:
\begin{align}
 S[g_{\mu\nu},A_\mu,\varphi]&=\frac{1}{16\pi}\int d^4x \sqrt{-g}\bigg(\! R-2g^{\mu\nu}\partial_\mu\varphi\partial_\nu\varphi-e^{-2a\varphi}F^{\mu\nu}F_{\mu\nu}\bigg)+S_{\rm m}[\Psi,\mathcal A^2(\varphi)g_{\mu\nu},A_\mu]\ ,\label{actionFondamEMD}
 \end{align}
 where $R$ is the Ricci scalar associated to $g_{\mu\nu}$, where $g=\det g_{\mu\nu}$, and  where $F_{\mu\nu}=\partial_\mu A_\nu-\partial_\nu A_\mu$. As for matter fields $\Psi$, they are minimally coupled to the Jordan metric $\tilde g_{\mu\nu}=\mathcal A^2(\varphi)g_{\mu\nu}$, $\mathcal A(\varphi)$ being a scalar function that specifies the theory, together with the coupling parameter $a$.\\
 
 In paper \cite{Julie:2017rpw}, we studied the conservative sector of the dynamics of compact binary systems in EMD theories. To do so, we phenomenologically replaced $S_{\rm m}$ in (\ref{actionFondamEMD}) by a point particle action which generalizes that of Eardley in scalar-tensor theories \cite{Eardley},
 \begin{align}
 S_{\rm m}\to S_{\rm m}^{\rm pp}[g_{\mu\nu},A_\mu,\varphi,\{x_A^\mu\}]=-\sum_A\int m_A(\varphi)\,ds_A+\sum_A q_A\int A_\mu\, dx^\mu_A\ ,\label{skeletonReplacement}
 \end{align}
where $ds_A=\sqrt{-g_{\mu\nu} dx_A^\mu dx_A^\nu}$, $x_A^\mu[s_A]$ being the worldline of body $A$, which is characterized by a constant charge $q_A$ and a ``sensitivity" $m_A(\varphi)$ that depends on its internal structure and on the value of the scalar field at its location.

The field equations derived from the ``skeleton" action (\ref{actionFondamEMD},\ref{skeletonReplacement}) read:
\begin{subequations}
\begin{align}
&R^{\mu\nu}-\frac{1}{2}g^{\mu\nu}R=8\pi \left(T^{\mu\nu}_{(\varphi)}+T^{\mu\nu}_{(A)}+T^{\mu\nu}_{({\rm m})}\right)\ ,\\
&\nabla_\nu\left(e^{-2a\varphi}F^{\mu\nu}\right)=4\pi\sum_A q_A\int\! ds_A\,\frac{\delta^{(4)}\left(x-x_A(s_A)\right)}{\sqrt{-g}}\,\frac{dx_A^\mu}{ds_A}\ ,\label{maxwellEqn}\\
&\Box\,\varphi=-\frac{a}{2}e^{-2a\varphi}F^2+4\pi\sum_A\int\! ds_A\,\frac{dm_A}{d\varphi}\frac{\delta^{(4)}\left(x-x_A(s_A)\right)}{\sqrt{-g}}\ ,\label{KGeqn}
\end{align}\label{twoBody_fieldEqn}%
\end{subequations}
where $\nabla_\mu$ denotes the covariant derivative, where $\delta^{(4)}\left(x-y\right)$ is the 4-dimensional Dirac distribution, and where
\begin{align}
T^{\mu\nu}_{(\varphi)}=\frac{1}{4\pi}&\left(\partial^\mu\varphi\,\partial^\nu\varphi-\frac{1}{2}g^{\mu\nu}(\partial\varphi)^2\right)\ ,\quad T^{\mu\nu}_{(A)}=\frac{1}{4\pi}e^{-2a\varphi}\left(F^{\mu\lambda} F^\nu_{\ \lambda}-\frac{1}{4}g^{\mu\nu}F^2\right)\ ,\label{stressEnergyTensors}\\
&\text{and}\quad T_{\rm (m)}^{\mu\nu}=\sum_A\int\! ds_A\, m_A(\varphi)\frac{\delta^{(4)}\left(x-x_A(s_A)\right)}{\sqrt{-g}}\frac{dx_A^\mu}{ds_A}\frac{dx_A^\nu}{ds_A}\ .\nonumber
\end{align}

In \cite{Julie:2017rpw}, we solved the field equations (\ref{twoBody_fieldEqn},\ref{stressEnergyTensors}) perturbatively around a flat, Minkowski background $\eta_{\mu\nu}$ and the constant value $\varphi_0$ of the scalar field at infinity, which is imposed by the cosmological environment of the binary system. We then derived the two-body Lagrangian, at post-keplerian order (1PK) and in harmonic coordinates, which generalizes that of Einstein, Hilbert and Hoffman in general relativity. To do so, we proceded \`a la Fichtenholz, which, at this order, is strictly equivalent to computing, e.g., a Fokker Lagrangian. The resulting Lagrangian depends on the quantities:
 \begin{align}
  \alpha_A^0 =\frac{d\ln m_A}{d\varphi}(\varphi_0)\ ,\quad \beta_A^0=\frac{d\alpha_A}{d\varphi}(\varphi_0)\ ,\quad\text{and}\quad  e_A=\frac{q_A}{m_A^0}e^{a\varphi_0}\ ,\label{bodyDepParameters}
 \end{align}
 and their $B$ counterparts, where and from now on, a 0 index indicates a quantity evaluated at infinity, $\varphi=\varphi_0$. We recall here its expression, introducing $R=\vert \vec x_A-\vec x_B\vert$, $\vec N=(\vec x_A-\vec x_B)/R$, and $\vec V_A= d\vec x_A/dt$:
\begin{align}
L&=-m_A^0-m_B^0+\frac{1}{2}m_A^0V_A^2+\frac{1}{2}m_B^0V_B^2+\frac{G_{AB}m_A^0 m_B^0}{R}\label{Lag1PK_PPN}\\
&+\frac{1}{8}m_A^0V_A^4+\frac{1}{8}m_B^0V_B^4+\frac{G_{AB}m_A^0 m_B^0}{R}\left[\frac{3}{2}(V_A^2+V_B^2)-\frac{7}{2}(V_A.V_B)-\frac{1}{2}(N.V_A)(N.V_B)+\bar \gamma_{AB}(\vec V_A-\vec V_B)^2\right]\nonumber\\
&-\frac{G_{AB}^2 m_A^0 m_B^0}{2R^2}\left[m_A^0(1+2\bar\beta_B)+m_B^0(1+2\bar\beta_A)\right]+\mathcal O(V^6)\ ,\nonumber
\end{align}
where $G_{AB}$, $\bar\gamma_{AB}$ and $\bar\beta_{A/B}$ are the following combinations of the body-dependent quantities (\ref{bodyDepParameters}):
\begin{subequations}
\begin{align}
&G_{AB}= G_*\left(1+\alpha_A^0\alpha_B^0-e_Ae_B\right)\ ,\label{gAB}\\ 
&\bar\gamma_{AB}=\frac{-4\,\alpha_A^0\alpha_B^0+3\,e_A e_B}{2(1+\alpha_A^0\alpha_B^0-e_Ae_B)}\ ,\label{PPNgamma}\\
&\bar\beta_A=\frac{1}{2}\frac{\beta_A^0{\alpha_B^0}^2-2\,e_Ae_B(a\,\alpha_B^0-\alpha_A^0\alpha_B^0)+e_B^2(1+a\,\alpha_A^0-e_A^2)}{(1+\alpha_A^0\alpha_B^0-e_Ae_B)^2}\quad\text{and}\quad A\leftrightarrow B\ ,\label{PPNbeta}
\end{align}\label{PPN}%
\end{subequations}
$G_*$ being Newton's constant in the Einstein frame, which we shall keep track of in the following for clarity.

From this Lagrangian, one can derive the linear momenta $P_A^i=\partial L/\partial V_A^i$ and the associated mechanical energy $E=P_A\cdot V_A+P_B\cdot V_B-L$.
For circular orbits, which is the case of interest, and in the center-of-mass frame (such that $P_A^i+P_B^i=0$), the 1PK mechanical energy $E$ can be expressed in terms of the orbital angular velocity $\dot\phi$ alone, using the equations of motion deduced from (\ref{Lag1PK_PPN}), and reads:
\begin{equation}
E=-\frac{1}{2}\mu\left(G_{AB}M\dot\phi\right)^{2/3}\left[1-\left(G_{AB}M\dot\phi\right)^{2/3}\left(\frac{3}{4}+\frac{\nu}{12}+\frac{2}{3}(\bar\gamma_{AB}-\langle\bar\beta\rangle)\right)+\mathcal O(V^4)\right]\ ,\label{mechanicalEnergy}
\end{equation}
where $M=m_A^0+m_B^0$, $\mu=m_A^0m_B^0/M$, and $\nu=\mu/M$, and where $\langle\bar\beta\rangle=\left(m_A^0\bar\beta_B+m_B^0\bar\beta_A\right)/M$.\\

In \cite{Julie:2017rpw}, we then particularized our results to binary systems composed of two charged, non-spinning black holes with vector and scalar ``hair", see Gibbons and Maeda \cite{Gibbons:1982ih, Gibbons:1987ps}. The sensitivity $m_A(\varphi)$ characterizing them was obtained analytically in \cite{Julie:2017rpw} for all $a$; in the simple case $a=1$ that we will consider here, it is given by:
\begin{equation}
m_A(\varphi)=\sqrt{\mu_A^2+q_A^2\frac{e^{2\varphi}}{2}}\ .
\end{equation}
Here $q_A$ is the constant $U(1)$ charge of the black hole appearing in the action (\ref{skeletonReplacement}). As for the constant $\mu_A$, it is its irreducible mass: $\mu_A=M_{\rm irr}$ (and not its ADM mass which is not conserved when orbiting aroud a companion). Since $M_{\rm irr}=\sqrt{S/4\pi}$, the constancy of $\mu_A$ implies that of the black hole entropy. This fact derives from the ``skeletonization" approximation, hence showing its limitations \cite{Cardenas:2017chu}.

The ``sensitivities" being known, all the body-dependent quantities (\ref{bodyDepParameters}) are known for a given black hole $A$, that is, for given values of ($q_A$, $\mu_A$), as functions of $\varphi_0$. In particular, as highlighted in \cite{Julie:2017rpw}, we have that $\alpha_A^0\equiv\alpha_A(\varphi_0)$ (which is an exact ``Fermi-Dirac distribution" when $a=1$) transitions from zero (Schwarzschild limit, with $\beta_A^0\to 0$ and $e_A^2\to 0$) to $a$ (fully scalarized black hole, with $\beta_A^0\to 0$ and $e_A^2\to 1+a^2$) when the scalar cosmological background $\varphi_0$ increases.\\


These previously obtained results being recalled, we now proceed to calculate the gravitational waves emitted by EMD compact binary systems. In order to describe the shrinking of the orbit due to gravitational radiation, we first compute the energy flux radiated away by the system:
\begin{equation}
-\frac{d\mathcal E}{dt}=\mathcal F_g+\mathcal F_A+\mathcal F_\varphi\ ,\label{energyBalance}
\end{equation}
\vspace*{-0.5cm}
\begin{subequations}
\begin{align}
\text{where}& \quad \mathcal E= \int d^3x\, |g|\left(T^{00}_{(\varphi)}+T^{00}_{(A)}+T^{00}_{({\rm m})}+t^{00}_{\rm LL}\right)\ ,\label{calE}\\
\mathcal F_g=\int_{x\to\infty}\!\! \!\!\!|g|\,t^{0i}_{{\rm LL}}\, n_i x^2 d\Omega^2\ &,\qquad \mathcal F_A=\int_{x\to\infty}\!\! \!\!\!|g|\,T^{0i}_{(A)}\, n_i x^2 d\Omega^2\ ,\qquad \mathcal F_\varphi=\int_{x\to\infty}\!\! \!\!\!|g|\,T^{0i}_{(\varphi)}\, n_i x^2 d\Omega^2\ ,\label{fluxCHflux}
\end{align}
\end{subequations}
with $n^i= x^i/x$, $x^i$ being the distance of the observation point from the source, and $d\Omega^2=\sin\theta\, d\theta d\phi$. $\mathcal F_g$ is the well-known flux in general relativity, $t^{\mu\nu}_{{\rm LL}}$ being the Landau-Lifshitz pseudo-tensor \cite{landau_classical_1975}, while $\mathcal F_A$ and $F_\varphi$ are the extra ``graviphotonic" and scalar fluxes.\\

The calculation of the fluxes (\ref{fluxCHflux}) is standard but a bit heavy, and will be detailed elsewhere (it is an extension of the general relativistic text-book calculation, see \cite{LeLivre}). The result, which is presented in \cite{maThese}, can be decomposed as follows:

The metric flux reduces, at leading order (that is 0PK), to that of Einstein's second quadrupole formula, but dressed up by the scalar and ``graviphotonic" contributions; that is, for circular orbits and in the center-of-mass frame:
\begin{equation}
\mathcal F_g=\frac{32}{5}\frac{\nu^2\left(G_{AB}M\dot\phi\right)^{10/3}}{G_*\left(1+\alpha_A^0\alpha_B^0-e_Ae_B\right)^2}+\cdots\ ,\label{fluxMetrique_CHAPflux}
\end{equation}
where we recall that $\dot\phi=d\phi/dt$ is the (gauge invariant) orbital angular velocity.

The ``graviphotonic" flux is dominated by a dipolar (-1PK) term. In order to determine its next-to-leading (0PK) contributions, one has to iterate (\ref{maxwellEqn}) to take into account the couplings to the metric and scalar field. We found:
\begin{align}
&\quad\mathcal F_A=\frac{\nu^2\left(G_{AB}M\dot\phi\right)^{8/3}}{G_*\left(1+\alpha_A^0\alpha_B^0-e_Ae_B\right)^2}\left\{\frac{2}{3}(e_A-e_B)^2\right.\label{fluxVecteur_CHAPflux}\\
&\qquad\qquad\qquad +\left(G_{AB}M\dot\phi\right)^{2/3}\left[\frac{8}{5}\left(\frac{m_A^0e_B+m_B^0e_A}{M}\right)^2+\frac{4}{9}(e_A-e_B)^2\Big(\nu-3-\bar\gamma_{AB}-2\langle\bar\beta\rangle\Big)\right.\nonumber\\
&\qquad\qquad\qquad\qquad +4(e_A-e_B)\left(\frac{(m_A^0)^2e_B-(m_B^0)^2e_A}{15M^2}\left.\left.-\frac{e_A(1+a\,\alpha_B^0)-e_B(1+a\,\alpha_A^0)}{3M(1+\alpha_A^0\alpha_B^0-e_Ae_B)}\right)\right]+\cdots\right\}\ ,\nonumber
\end{align}
where $\bar\gamma_{AB}$ and $\langle\bar\beta\rangle$ are given in (\ref{PPN}) and seq. Note that when the scalar field is switched off, this formula gives the electric flux emitted by charged systems such as Reissner-Nordstr\"om binary black holes (which, in itself, is also a new result at this order). Note also that when $e_A=e_B$, the dipolar contribution disappears and $\mathcal F_A$ identifies to the maxwellian quadrupolar flux,
 the charges being dressed up by the value of the background scalar field. 

Finally, the scalar flux is \textit{a priori} monopolar (-2PK), see, e.g., \cite{Damour:1992we}. However, for circular orbits to which we restrict ourselves here, it is dipolar (-1PK), and its 0PK part is obtained by iterating (\ref{KGeqn}), to take into account its coupling to the metric and to the ``graviphoton". It is given by:
\begin{align}
&\mathcal F_\varphi=\frac{\nu^2\left(G_{AB}M\dot\phi\right)^{8/3}}{G_*\left(1+\alpha_A^0\alpha_B^0-e_Ae_B\right)^2}\Bigg\{\frac{1}{3}(\alpha_A^0-\alpha_B^0)^2\label{fluxScalaire_CHAPflux}\\
&\quad\quad+\left(G_{AB}M\dot\phi\right)^{2/3}\left[\frac{16}{15}\left(\frac{m_A^0\alpha_B^0+m_B^0\alpha_A^0}{M}\right)^2+\frac{2}{9}(\alpha_A^0-\alpha_B^0)^2\Big(\nu-3-\bar\gamma_{AB}-2\langle\bar\beta\rangle\Big)\right.\nonumber\\
&\quad\quad+2(\alpha_A^0-\alpha_B^0)\left(\frac{(m_A^0)^2\alpha_B^0-(m_B^0)^2\alpha_A^0}{5M^2}
\left.\left.+\frac{m_A^0\left[\alpha_B^0+\alpha_A^0\,(\alpha_B^0)^2+\beta_B^0\alpha_A^0-a\,e_Ae_B\right]-(A\leftrightarrow B)}{3M(1+\alpha_A^0\alpha_B^0-e_Ae_B)}\right)\right]+\cdots\right\}\nonumber\ .
\end{align}
When $\alpha_A^0=\alpha_B^0$, this flux is reduced to its purely (scalar) quadrupolar term. This expression generalizes that of Damour and Esposito-Far\`ese in scalar-tensor theories \cite{Damour:1992we}, which is recovered when $e_{A/B}=0$; note however that the EMD flux cannot be deduced from \cite{Damour:1992we} by a mere generalization of $G_{AB}$ to include $e_{A/B}$, due to the presence of $a$ in the last term of (\ref{fluxScalaire_CHAPflux}).\\

From these fluxes, we can then determine the characteristics of the ``chirp", i.e., of the evolution $\ddot\phi$ of the orbital velocity $\dot\phi$ at 0PK order.
To do so, one assumes that the energy $\mathcal E$, given in (\ref{calE}) and whose radiative decay is given in (\ref{energyBalance}), is equal to the mechanical energy of the system, $E$, given in (\ref{mechanicalEnergy}). This yields:

\begin{align}
\ddot\phi=&G_*\mu\,\dot\phi^3\Bigg\{\bigg[2(e_A-e_B)^2 +(\alpha_A^0-\alpha_B^0)^2\bigg]\label{balance_CHAPflux}\\
&+\left(G_{AB}M\dot\phi\right)^{2/3}\bigg[\frac{96}{5}+\frac{24}{5}\left(\frac{m_A^0e_B+m_B^0e_A}{M}\right)^2\!+\frac{16}{5}\left(\frac{m_A^0\alpha_B^0+m_B^0\alpha_A^0}{M}\right)^2\nonumber\\
&\qquad\qquad\qquad\nonumber+\bigg(2(e_A-e_B)^2+(\alpha_A^0-\alpha_B^0)^2\bigg)\left(\frac{5\nu}{6}-\frac{1}{2}+\frac{2}{3}\bar\gamma_{AB}-\frac{8}{3}\langle\bar\beta\rangle\right)\\
&\qquad\qquad\qquad +4(e_A-e_B)\left(\frac{(m_A^0)^2e_B-(m_B^0)^2e_A}{5M^2}-\frac{e_A(1+a\,\alpha_B^0)-e_B(1+a\,\alpha_A^0)}{M(1+\alpha_A^0\alpha_B^0-e_Ae_B)}\right)\nonumber\\
&+2(\alpha_A^0-\alpha_B^0)\left(\frac{3}{5}\frac{(m_A^0)^2\alpha_B^0-(m_B^0)^2\alpha_A^0}{M^2}
+\frac{m_A^0\left[\alpha_B^0+\alpha_A^0\,(\alpha_B^0)^2+\beta_B^0\alpha_A^0-a\,e_Ae_B\right]-(A\leftrightarrow B)}{M(1+\alpha_A^0\alpha_B^0-e_Ae_B)}\right)\bigg]+\cdots\Bigg\}\ .\nonumber
\end{align}

In the general relativistic limit ($e_{A/B}=0$, $m_{A/B}(\varphi)=cst$), equation (\ref{balance_CHAPflux}) is reduced to the well-known expression $\ddot\phi\propto\dot\phi^{11/3}$:
\begin{equation}
\ddot\phi=\frac{96}{5}(G_*\mathcal M)^{5/3}\dot\phi^{11/3}\ ,\quad\text{where}\quad \mathcal M=\nu^{3/5}M\label{chirpMassRG_CHAPflux}
\end{equation}
is the ``chirp mass".
By contrast, in presence of dipolar radiation, the right-hand side of (\ref{balance_CHAPflux}) is dominated, during early inspiral, by its first term, $\ddot\phi\propto\dot\phi^{3}$:
\begin{equation}
\ddot\phi=(G_*\mathcal M_{\mathcal D}) \dot\phi^3\quad\text{with}\quad \mathcal M_{\mathcal D}=\nu\bigg[2(e_A-e_B)^2+(\alpha_A^0-\alpha_B^0)^2\bigg]M\ ,\label{chirpMassDipole_CHAPflux}
\end{equation}
which governs the motion of a binary system when, for example, one of its components is a scalarized EMD black hole described above. Finally, when the dipoles are negligible ($\alpha_A^0\simeq\alpha_B^0$ and $e_A\simeq e_B$), we have again $\ddot\phi\propto\dot\phi^{11/3}$, but the general relativistic ``chirp mass" is now dressed up by the vector and scalar quadrupoles:
\begin{align}
&\hspace*{5cm}\ddot\phi=\frac{96}{5}(G_*\mathcal M_{\mathcal Q})^{5/3}\dot\phi^{11/3}\quad\text{with}\label{chirpMassQuadrupole_CHAPflux}\\
&\mathcal M_{\mathcal Q}=\nu^{3/5}(1+\alpha_A^0\alpha_B^0-e_Ae_B)^{2/5}\left[1+\frac{1}{4}\left(\frac{m_A^0e_B+m_B^0e_A}{M}\right)^2+\frac{1}{6}\left(\frac{m_A^0\alpha_B^0+m_B^0\alpha_A^0}{M}\right)^2\right]^{3/5}\!\!\! M\ .\nonumber
\end{align}

As we shall see below, the angular velocity $\dot\phi$ is proportional to the frequency of the gravitational wave observed at infinity, and is hence a combination that can be measured numerically.
The results above therefore show that in presence of significant dipolar radiation, cf. (\ref{chirpMassDipole_CHAPflux}), the evolution of the frequency $f$ will deviate from the general relativistic predictions. In contrast, when dipolar radiation is absent, the ``chirp" is reduced to (\ref{chirpMassQuadrupole_CHAPflux}), and the deviations from general relativity can, at this order, be absorbed in a redefinition of the masses.
\\


Having in hand the fluxes (\ref{fluxMetrique_CHAPflux}-\ref{fluxScalaire_CHAPflux}), one can also deduce the radiation reaction force exerted on the system by equating its power to the energy fluxes at infinity \cite{Buonanno:2000ef}. For the quasi-circular orbits considered here, this force is tangent to the trajectory and reads
\begin{equation}
F_\phi=-\frac{\mathcal F_g+\mathcal F_A+\mathcal F_\varphi}{\dot\phi}\ .\label{radiationForce}
\end{equation}
The equations of motion, deduced from the 1PK Lagrangian computed in \cite{Julie:2017rpw} and recalled above, can hence be generalized to encompass radiation reaction effects as:
\begin{equation}
\frac{d}{dt}\left(\frac{\partial L_{1PK}}{\partial \dot R}\right)-\frac{\partial L_{1PK}}{\partial R}=0\ ,\quad \frac{d}{dt}\left(\frac{\partial L_{1PK}}{\partial \dot \phi}\right)-\frac{\partial L_{1PK}}{\partial \phi}=F_\phi\ .\label{eomPK}
\end{equation}
Now, the range of validity of these equations of motion can be extended, hopefully up to merger, by resumming them. To do so, we shall start, not from the 1PK Lagrangian, but rather, from the scalar-tensor effective-one-body (EOB) Hamiltonian presented in \cite{Julie:2017pkb}: indeed, it is trivially generalized to EMD theories at 1PK order, as we showed in \cite{Julie:2017rpw}, since the Lagrangian (\ref{Lag1PK_PPN}) has exactly the same structure as that of scalar-tensor theories at this order. In this EOB approach, the equations of motion (\ref{eomPK}) are replaced by
\begin{align}
&\dot r=\frac{\partial H_{\rm EOB}}{\partial p_r}\quad, \quad \dot p_r=-\frac{\partial H_{\rm EOB}}{\partial r}\quad ,\quad \dot \phi=\frac{\partial H_{\rm EOB}}{\partial p_\phi}\quad, \quad \dot p_\phi=-\frac{\partial H_{\rm EOB}}{\partial \phi}+F_\phi\ ,\label{EOBeom_CHAPflux}\\
&\text{with}\quad H_{\rm EOB}=M\sqrt{1+2\nu\left(\frac{H_e}{\mu}-1\right)}\ ,\quad\text{where}\quad H_e=\mu\sqrt{A\left(1+\frac{p_r^2}{\mu^2 B}+\frac{p_\phi^2}{\mu^2 r^2}\right)}\ ,\label{HamiltonienEOB_CHAPeobst}
\end{align}
\vspace*{-0.5cm}
\begin{subequations}
\begin{align}
 \text{and}\quad & A(r)=\mathcal P^1_1\left[ 1-2\left(\frac{G_{AB}M}{r}\right)+2\bigg[\langle\bar\beta\rangle-\bar\gamma_{AB}\bigg]\left(\frac{G_{AB}M}{r}\right)^2\,\right]\ ,\\
 &B(r)=1+2 \bigg[1 + \bar\gamma_{AB}\bigg]\left(\frac{G_{AB}M}{r}\right)\ ,
 \end{align}
 \end{subequations}
where $\mathcal P^1_1$ denotes the Pad\'e approximant of order $(1,1)$, with respect to the variable $u=(G_{AB}M/r)$, and where $(r,\phi; p_r, p_\phi)$ are the effective phase space coordinates introduced in \cite{Julie:2017pkb}, such that $\phi$ identifies, for circular orbits, to the (observable) orbital phase used above. When considered as exact, these equations can be numerically integrated to yield the evolution of the binary system up to the innermost stable circular orbit $u_{\rm ISCO}$, which is defined as the (outermost) solution of $A''/A'=(Au^2)''/(Au^2)'$ (for a detailed study of the ISCO and its characteristics, including, e.g., the associated orbital frequency in scalar-tensor and EMD theories, see, again, \cite{Julie:2017pkb}). As for the initial conditions, they are determined exactly as in general relativity, see \cite{Buonanno:2000ef}.

Figure \ref{FigureTrajectoire_CHAPflux} below shows the illustrative example of two EMD black holes for the theory $a=1$, the first one being scalarized, while the second one is a Schwarzschild black hole. As expected, a strong dipolar (i.e. ``graviphotonic" and scalar) radiation is driving the radius of the orbit to decrease at a much greater rate than in general relativity.\\

\begin{figure}[!h]
\centering
 \includegraphics[width=0.45\linewidth]{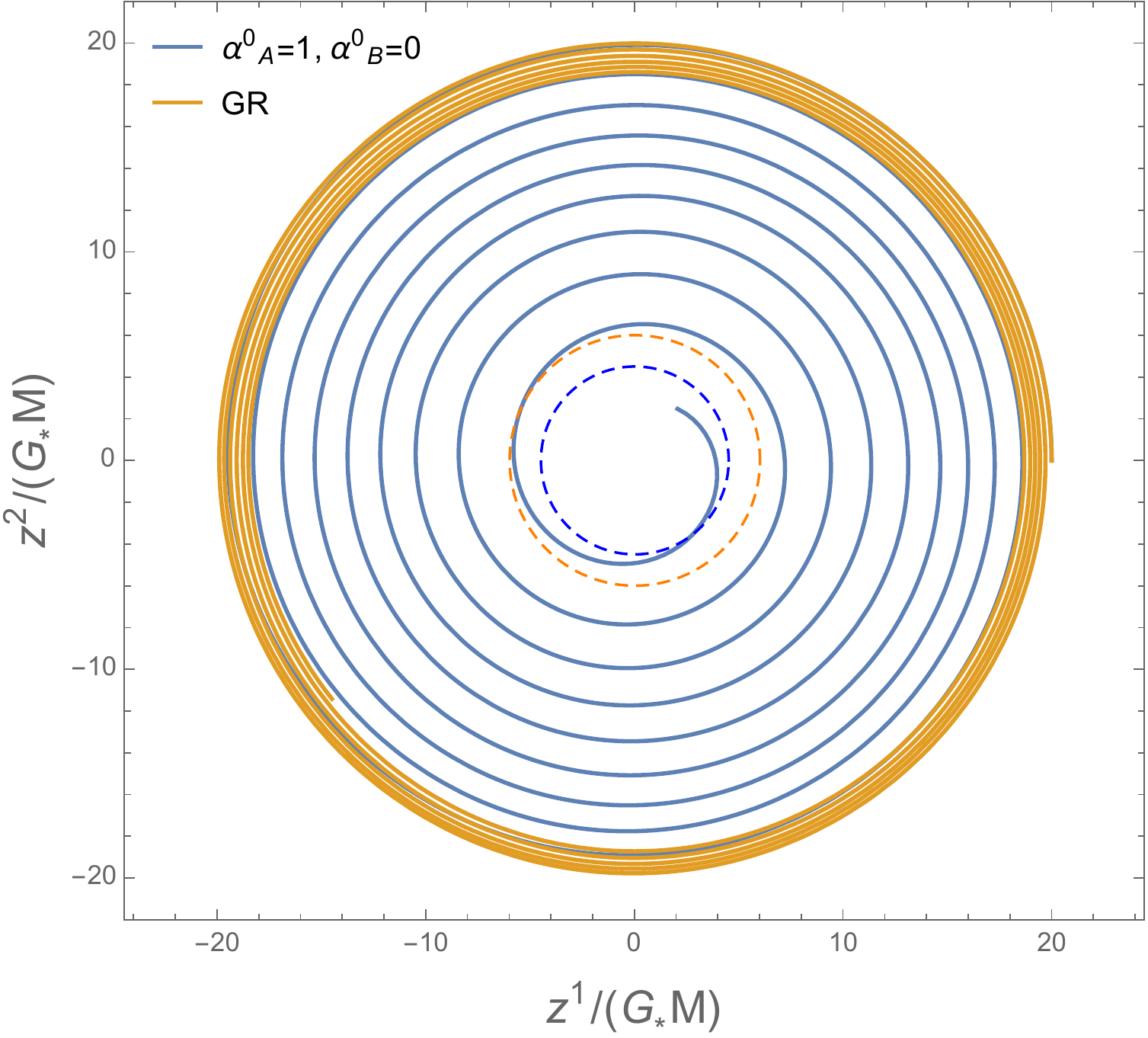}
 \caption{Effective trajectories ($z^1=r\cos\phi$, $z^2=r\sin\phi$) for a binary system composed of two EMD black holes with identical masses ($\nu=1/4$), in the theory $a=1$. In blue, one of the black holes is scalarized ($\alpha_A^0=1$, $e_A=\sqrt{2}$), while the second one is Schwarzschild's ($\alpha_B^0=0$, $e_B=0$); in orange, the general relativistic case of two Schwarzschild black holes. The corresponding ISCOs are also shown (dashed lines). Starting from $r=20\,G_*M$, the scalarized system reaches its ISCO within $\Delta t\simeq 3000\,G_*M$, to be compared to $\Delta t\simeq10\,000\,G_*M$ in general relativity. Note that we have $(\dot r/r\dot\phi)^2_{\rm ISCO}=0.019$, hence justifying the validity of our quasi-circular approximation, at least up to the ISCO.}
 \label{FigureTrajectoire_CHAPflux}
\end{figure}

The trajectories being known and pushed up to the ISCO, one easily predicts, at leading order, the associated waveforms.

Although we can compute the vector component of the waveform, this is not necessary, supposing that the detectors are ``gravielectrically" neutral. Rather, the mirrors of the interferometers follow the geodesics of the Jordan metric, see (\ref{actionFondamEMD}) and seq., which reads, in the solar system:
\begin{equation}
\tilde g_{\mu\nu}=\mathcal A^2(\varphi)g_{\mu\nu}=\mathcal A^2_\odot\left[\eta_{\mu\nu}(1+2\alpha_\odot\delta\varphi)+h_{\mu\nu}^{TT}\right]+\mathcal O\left(\frac{1}{x^2}\right)\ ,\label{jordanMetric_CHAPflux}
\end{equation}
where $x$ is the distance to the source, $\mathcal A_\odot=\mathcal A(\varphi_\odot)$ is the value of the coupling of matter to the scalar field in the solar system, and where $\alpha_\odot=(d\ln\mathcal A/d\varphi)(\varphi_\odot)$. As for $\delta\varphi=\varphi-\varphi_\odot$, it is the scalar wave, given at leading order by
\begin{equation}
\delta\varphi=-G_*\frac{n_i\dot{\mathcal D_S^{\,i}}}{x}\quad\text{with}\quad\mathcal D_S^i=\sum_A m_A^0\alpha_A^0 x_A^i\ ,\label{ondeScalaire_CHAPflux}
\end{equation}
$n_i\dot{\mathcal D_S^{\,i}}$ being the projection of the time derivative of the (scalar) dipole $\mathcal D_S^i$ of the source on the line of sight $n^i=x^i/x$, and where $x_{A/B}^i$ denote the (harmonic) position of the bodies in their center-of-mass frame. Finally, the useful components of $h_{\mu\nu}^{TT}$ are
\begin{equation}
h_{ij}^{TT}=\frac{2G_*}{3}\frac{\mathcal P_{ij}^{\ kl}\ddot{\mathcal Q}_{kl}}{x}\quad\text{where}\quad \mathcal Q^{ij}=\sum_A m_A^0\left(3x_A^ix_A^j-\delta^{ij}x_A^2\right)\ ,
\end{equation}
$\mathcal P_{ij}^{\ kl}\ddot{\mathcal Q}_{kl}$ being the transverse (to $n^i$) and traceless part of the second time derivative of the (mass) quadrupole of the system. Note also that although the new scalar polarization (\ref{ondeScalaire_CHAPflux}) cannot be disentangled from the tensor ones for the time being, it will when a third detector (e.g., KAGRA) joins the LIGO-Virgo network, see, e.g., \cite{Abbott:2017oio}. \\

We note that $\delta\varphi$ is of -0.5PK order relative to $h_{ij}^{TT}$. However, the contribution of $\delta\varphi$ to the wave (\ref{jordanMetric_CHAPflux}) is numerically lowered by the factor $\alpha_\odot$, which is already constrained by $\alpha_\odot\lesssim 10^{-2}$ from solar system observations, see, e.g., \cite{Bertotti:2003rm}. It is hence not necessary to compute the 0PK corrections to $\delta\varphi$. Let us however emphasize that the quantities $\alpha_{A/B}^0$ appearing in (\ref{ondeScalaire_CHAPflux}) are evaluated on the cosmological environment of the sources, $\varphi_0$; and, for scalarized black holes, they can numerically reach the order of unity.\\

\begin{figure}[h!]
\centering
\begin{subfigure}{.5\textwidth}
\hspace*{-1.9cm}
\includegraphics[width=10cm]{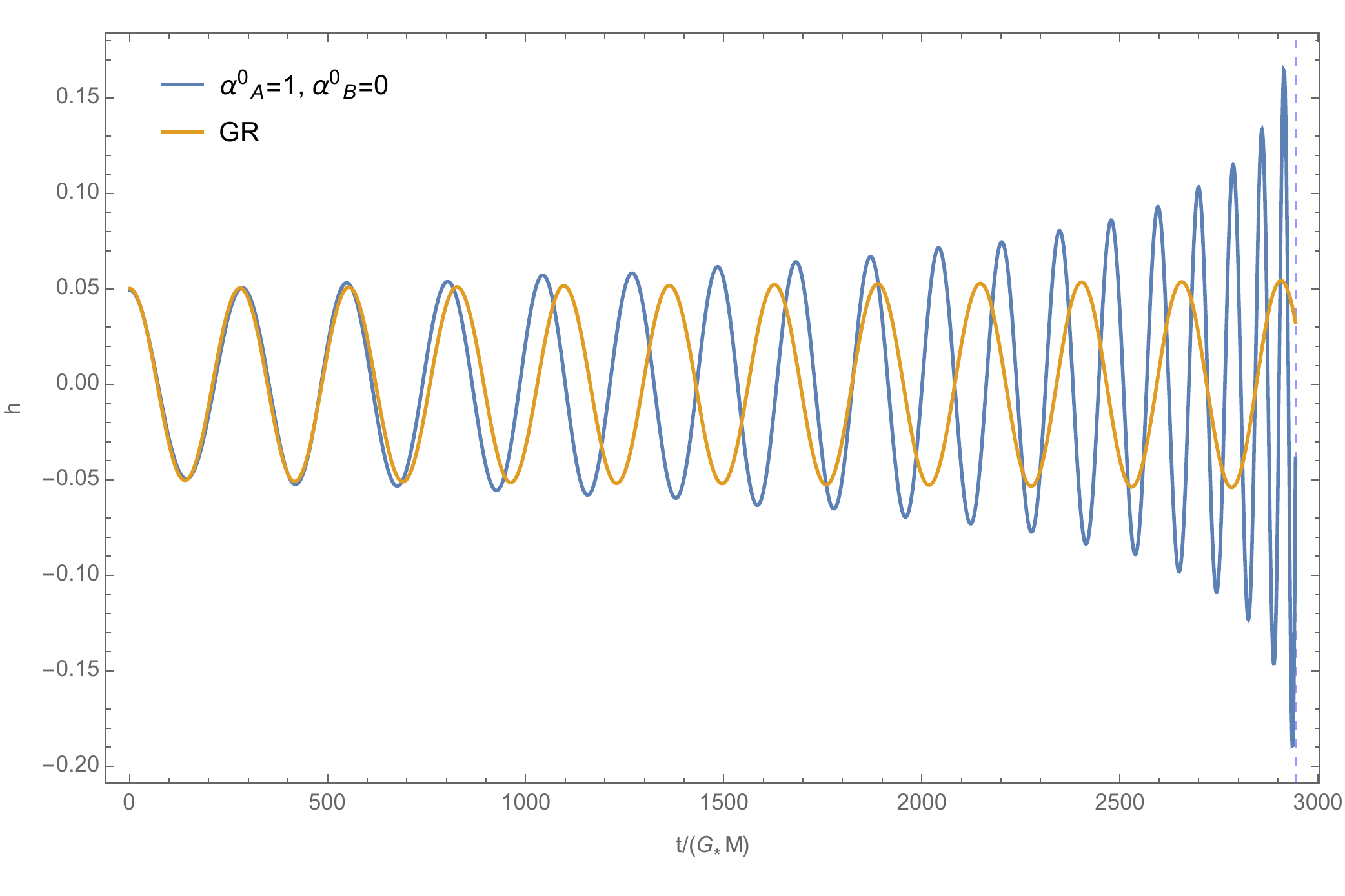}
\label{figureA}
\end{subfigure}%
\begin{subfigure}{.5\textwidth}
\hspace*{-0.3cm}
\includegraphics[width=10cm]{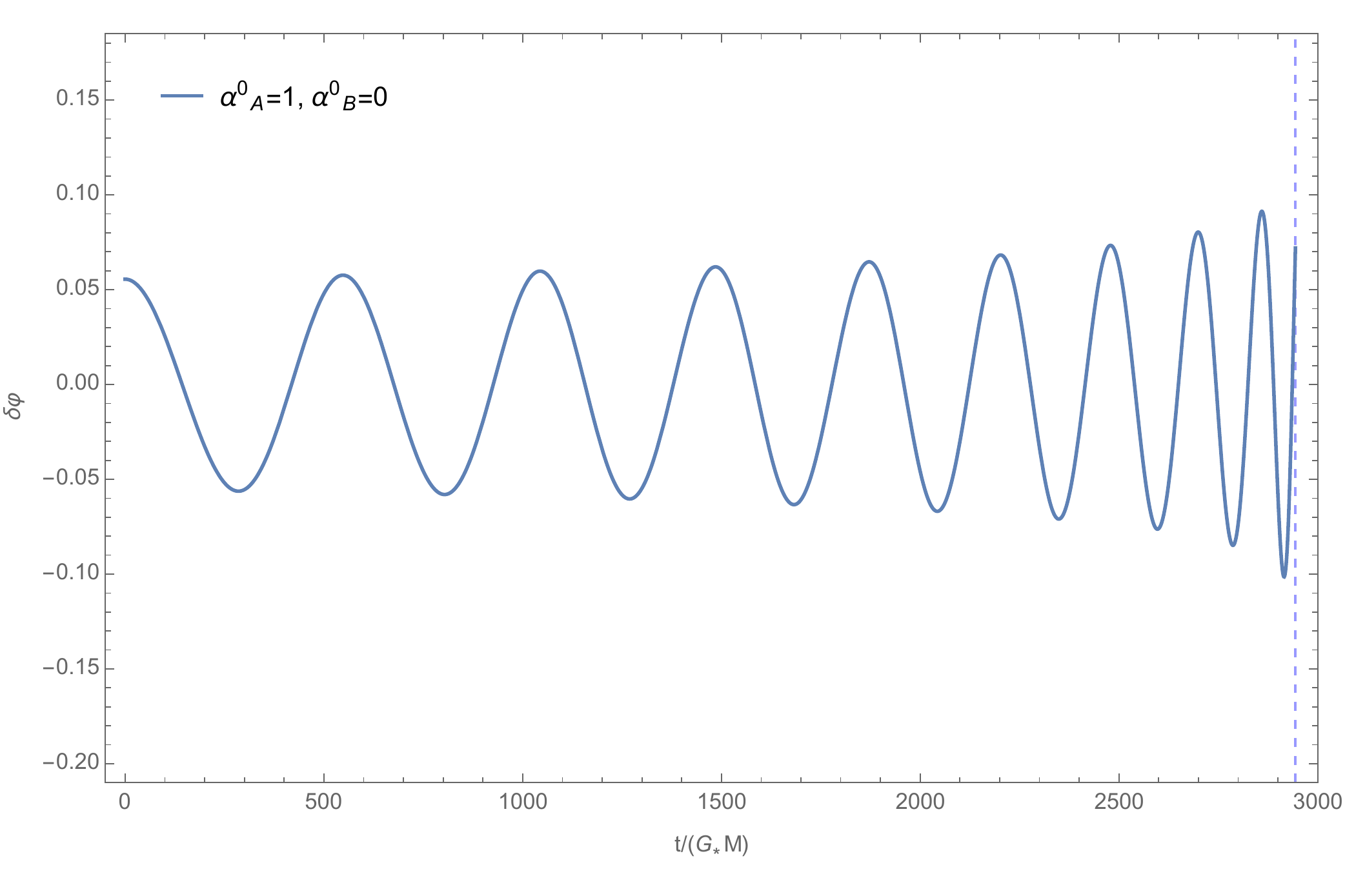}
\end{subfigure}%
\caption{Gravitational waveforms associated to the trajectories ploted in figure \ref{FigureTrajectoire_CHAPflux}, for the theory $a=1$. On the left panel, $h=\left(G_{AB}M\dot\phi\right)^{2/3}\cos(2\phi)$~; on the right panel, $\delta\varphi=(1/4)(\alpha_A^0-\alpha_B^0)\left(G_{AB}M\dot\phi\right)^{1/3}\cos(\phi)$. The scalar waveform amplitude is, in this example, numerically comparable to the tensor one; however its contribution to (\ref{jordanMetric_CHAPflux}) is numerically lowered by the solar system factor $\alpha_\odot$. The moment at which the system crosses the ISCO is represented by vertical dashed lines.}\label{FigureWaveform_CHAPflux}
\end{figure}

At the order considered here, $\delta\varphi$ and $h_{ij}^{TT}$ are computed using Kepler's laws for circular orbits, and can be expressed as functions of the angular velocity $\dot\phi$ only to yield:

\vfill\eject

\begin{subequations}
\begin{align}
&\delta\varphi=(\alpha_A^0-\alpha_B^0) \frac{G_*M\nu}{x}(\sin i)\left(G_{AB}M\dot\phi\right)^{1/3}\cos\phi\ ,\\
&h_+=\frac{4G_*M\nu}{x}\left(\frac{1+\cos^2i}{2}\right)\left(G_{AB}M\dot\phi\right)^{2/3}\cos(2\phi)\ ,\label{hplus_CHAPflux}\\
&h_\times=\frac{4G_*M\nu}{x}(\cos i)\left(G_{AB}M\dot\phi\right)^{2/3}\sin(2\phi)\ ,\label{hcroix_CHAPflux}
\end{align}\label{waveforms}%
\end{subequations}
where $i$ denotes the angle between the normal to the orbital plane and the line of sight.
The last step consists in injecting in (\ref{waveforms}) the trajectory $\phi(t)$ obtained when integrating (\ref{EOBeom_CHAPflux}). The waveforms $\delta\varphi$ and $h_{ij}^{TT}$ are thereby known up to the ISCO. Figure (\ref{FigureWaveform_CHAPflux}) gives those associated to the trajectories shown in figure \ref{FigureTrajectoire_CHAPflux}. \\

Finally, we note that for the Jordan metric (\ref{jordanMetric_CHAPflux}) to reduce to that of Minkowski in the absence of gravitational waves, one has to perform the local coordinate change $d\tilde t=\mathcal A_\odot dt$, $d\tilde x^i=\mathcal A_\odot dx^i$. Therefore, the observed frequency of the signal (\ref{hplus_CHAPflux},\ref{hcroix_CHAPflux}) in the solar system, defined by $2\pi f\equiv 2d\phi/d\tilde t$, is given by $f=\dot\phi/(\pi\mathcal A_\odot)$.


\section*{Ackowledgements}
I am very grateful to Nathalie Deruelle for her invaluable support and advice during the preparation of this manuscript; any shortcoming being, of course, mine.

\vspace*{0.5cm}

\noindent
\textit{Note added in proof:} The new results concerning the fluxes and waveforms presented in this paper represent part of my PhD thesis \cite{maThese}. While I was about to complete this paper, Khalil et al. submitted partly similar results in \cite{Khalil:2018aaj}.

\bibliographystyle{unsrt}

\end{document}